\documentclass[submission,copyright,creativecommons]{eptcs}
\providecommand{\event}{ICLP 2026} 

\usepackage{iftex}
\usepackage{graphicx}
\usepackage[mathscr]{euscript}
\usepackage{enumitem}

\ifpdf
  \usepackage{underscore}         
  \usepackage[T1]{fontenc}        
\else
  \usepackage{breakurl}           
\fi

\title{Explainability Framework for Policy-Aware Autonomous Agents}
\author{Heather Merhout
\institute{Miami University\\ Oxford, Ohio, United States}
\email{merhouhs@miamioh.edu}
\and
Daniela Inclezan
\institute{Miami University\\
Oxford, Ohio, United States}
\email{inclezd@miamioh.edu}
}
\def\titlerunning{Explainability Framework for Policy-Aware Autonomous Agents}
\def\authorrunning{Heather Merhout, D. Inclezan}
\begin{document}
\maketitle

\begin{abstract}
In the field of Artificial Intelligence, an \textit{agent} is a system which is able to autonomously make decisions in order to reach a desired goal.  As these systems grow more prevalent in our day-to-day lives, there has been an increased need to add explainability features which can provide an account for an agent's behavior.  We therefore propose a framework that outlines how to produce comprehensible explanations for policy-aware agents, or agents which have rule-enforcing policies incorporated in their decision-making framework.  This framework is designed using insights from the social sciences on how to produce good explanations. It is implemented in the Answer Set Programming language while using Python to assist with information extraction and natural-language translation.  Because these agents incur penalties when violating policies, we are able to leverage these penalties to detect undesirable events in scenarios that are counterfactual to the agents' original actions. This lends itself to creating contrastive explanations (e.g., ``the agent performed this action because, had it not, undesirable event X would have occurred.''), which formulate the core component for our explainability framework.  The framework is evaluated using a survey wherein human participants provide feedback on our program-generated explanations.
\end{abstract}

\section{Introduction}

As computers grow more and more prevalent in modern life and work, so too does the use of autonomous agents to make decisions based on a world of data. However, autonomous agents can sometimes act in unexpected or even undesirable ways.  Hence, our research seeks to promote ethical behavior in intelligent systems by proposing an explanation framework, which articulates the rationale behind an agent’s decision-making processes, for agents that are designed using Answer Set Programming \cite{gel91b}.

Explanation frameworks have been or are being developed for numerous different types of automated systems (see, for example, \cite{hayes:improving:2017} and \cite{hanna:application:2020}).  For this research, our explanation framework will apply primarily to planning and scheduling domain models which utilize policies within their decision-making framework. Because the goal is to explore methodologies and philosophies for constructing an explainability framework, the processes described in this paper may have applications in the broader field of AI as well. 


The motivation for this research stems from the increased desire for transparency across AI systems. As autonomous agents are being more and more integrated into systems such as self-driving cars and medical diagnostic systems, it is crucial that humans can trust these systems to make reasonable decisions based on the data they receive. An explanation system, wherein an agent can output a concise and comprehensible summary of the logical reasons for executing an action, will give a user insight into the functionality of an intelligent agent which it may employ for important tasks; not only does this give the user more confidence and trust that their agent is working within reasonable expectations, but it also\textemdash in the case that an agent does not work as its user desires\textemdash could serve to highlight where an agent's shortcomings exist and what changes must be made to the agent if possible.





\section{Background: Policies and $\mathscr{AOPL}$}
This research focuses on agents with policy awareness.  A policy-aware agent is one for which policies, or sets of rules/norms, are incorporated into its decision-making framework.  Of all the decisions an agent can possibly make, a policy is a rule defined by a subset of these possible trajectories which the agent's programmer deems preferable \cite{gelfond:authorization:2008}.

While simple rules, such as “if it is raining, do not perform action A,” can be implemented through Answer Set Prolog (ASP), it becomes increasingly important to describe a complicated set of rules to ensure an agent does not produce unwanted behavior.  One language for describing complex policies is known as \begin{math}\mathscr{AOPL}\end{math} (Authorization and Obligation Policy Language) \cite{gelfond:authorization:2008}, defined by Gelfond and Lobo. \begin{math}\mathscr{AOPL}\end{math} differs from both classical deontic logics and more recent deontic extensions of ASP in its design goals and level of abstraction. Classical deontic logic, originating with von Wright \cite{von:wright:i:1951} and developed in modal frameworks such as those surveyed by Hilpinen \cite{hilpinen:deontic:1971} and Meyer and Wieringa \cite{meyer:deontic:1993}, provides an axiomatic treatment of obligation, permission, and prohibition using modal operators, but remains largely static and not directly executable. In contrast, \begin{math}\mathscr{AOPL}\end{math} adopts the rule-based, nonmonotonic semantics of Answer Set Programming to encode norms as predicates over actions and states, enabling reasoning about authorization and obligation in dynamic domains with time, causality, and exceptions.  Cabalar’s work on deontic ASP \cite{balduccini:founded:2019, cabalar:deolingo:2025} takes a different route by extending the logical foundations of ASP with explicit deontic operators, yielding formalisms such as Deontic Equilibrium Logic where obligations and permissions are first-class constructs with precise semantics and can be compiled into ASP. Thus, while \begin{math}\mathscr{AOPL}\end{math} is primarily an application-oriented policy language that represents norms within an action theory, Cabalar’s approach is semantics-oriented, aiming to reconstruct deontic reasoning itself inside ASP, effectively bridging the gap between traditional deontic logic and nonmonotonic logic programming. Given that \begin{math}\mathscr{AOPL}\end{math} allows describing policies at a higher-level of abstraction than ASP, we adopted it for our work.

\subsection{Authorizations and Obligations}
\begin{math}\mathscr{AOPL}\end{math} distinguishes between policy authorizations and obligations. Authorizations (the ``\begin{math}\mathscr{A}\end{math}'' in \begin{math}\mathscr{AOPL}\end{math}) are policies that describe whether an agent is permitted (authorized) or not permitted to perform an action under certain circumstances.  An ``action'' in this context is classified as an \textit{elementary action $(e)$}, which consists of a single true or false atom, or a \textit{compound action $(c)$}, which consists of multiple simultaneous elementary actions.  An authorization is specified by using the predicate $permitted(e)$ to describe an action.  A strict authorization is one with no exceptions, while a \textit{defeasible} rule ($d$) is one for which exceptions are allowed. For a set of default authorization rules $d_1...d_n$, the predicate $prefer(d_1, d_2)$ is used to denote a preference of one authorization over another. 

Gelfond and Lobo describe authorization as expressions taking on one of these forms \cite{gelfond:authorization:2008}:

\begin{enumerate}[itemsep=0pt]
    \item $permitted(e)$ if $cond$
    \item $\neg permitted(e)$ if $cond$
    \item $d$ : normally $permitted(e)$ if $cond$
    \item $d$ : normally $\neg permitted(e)$ if $cond$
    \item $prefer(d_1, d_2)$
\end{enumerate}

It is important to note that a strict rule of type 2 above is not the same as an ASP constraint.  Rather than forbidding any world in which the policy is violated, a program will instead incur a penalty when a policy is violated, as will be seen later.

Obligations (the ``$\mathscr{O}$'' in \begin{math}\mathscr{AOPL}\end{math}) are policies which describe actions that an agent \textit{must} do or refrain from doing under certain circumstances.  Obligations are indicated by the predicate $obl(e)$ if an action $e$ is to be taken under certain conditions, or $obl(\neg e)$ if the agent is to refrain from action $e$ under certain conditions. Obligation expressions take on one of the following forms, for any $happening \in \{e, \neg e\}$ \cite{gelfond:authorization:2008}:

\begin{enumerate}[itemsep=0pt]
    \item $obl(happening)$ if $cond$
    \item $\neg obl(happening)$ if $cond$
    \item $d$ : normally $obl(happening)$ if $cond$
    \item $d$ : normally $\neg obl(happening)$ if $cond$
    \item $prefer(d_1, d_2)$
\end{enumerate}

\subsection{Rule Syntax}
When encoding a policy using $\mathscr{AOPL}$, the following syntax is often used (see \cite{tummala:penalization:2024}):

\begin{center}
    $r(t_1,...t_n)$: $head$ \textbf{if} $cond$
\end{center}
where $r$ is the label for the rule, $head$ is the head of an authorization or obligation, $cond$ is the condition (environmental state) which triggers the rule, and $t_1,...t_n$ is the set of $n$ terms which appear in the rule.  Examples of this syntax, which describe policies for a self-driving car, are as follows:

\begin{itemize}
\item $r1(I)$: normally $permitted(drive(I))$ \textbf{if} $light\_color(green, I)$
\item $r2(I)$: $obl(stop(I))$ \textbf{if} $pedestrians\_are\_crossing(I)$
\item $r3(I)$: $\neg permitted(drive(I))$ \textbf{if} $has\_sign(I, stop), \neg has\_speed(0).$
\end{itemize}

The rules can respectively be read as follows:
\begin{itemize}
\item Rule 1: the agent is normally permitted to drive through intersection I if the intersection's light color is green.
\item Rule 2: the agent is obligated to stop at intersection I if pedestrians are crossing the intersection.
\item Rule 3: the agent is not permitted to drive through intersection I if the intersection has a stop sign and the vehicle does not have a speed of 0 (i.e., it is not stopped).
\end{itemize}

Thus, when developing a planning framework for policy-aware autonomous agents, researchers can write rules that ensure a policy-compliant agent chooses a plan which not only achieves its desired goal but does so without violating policies implemented within its decision-making framework \cite{meyer2021apia,inclezan:ASP:2023, harders:plan:2023,InclezanHT24,TummalaI24,TUMMALA:INCLEZAN:2026}. However, there still remains a gap in knowledge regarding the development of an explainability framework, which intends to increase trustworthiness in these agents by providing a standardizable methodology to explain the plans that they take. The next section will outline a philosophical framework, adapted from \cite{miller:explanation:2019}, on what kind of information to include in an explanation for an autonomous agent.

\section{Explainability: A Philosophical Framework}
In 2019, Miller conducted a survey which collected information from various sources about what a human looks for in an explanation and how these findings are applicable to explaining AI programs \cite{miller:explanation:2019}.  It is this survey from which we will present the most relevant insights which can be used to build our framework for explaining our policy-aware autonomous agents.

\subsection{A Social Process}

One simple definition of an explanation that Miller cites, from Josephson and Josephson \cite{josephson:abductive:1996}, states,

\begin{quote}
``An explanation is an assignment of causal responsibility.''
\end{quote}

It is easy, especially when presented with such a definition, to simply think of an explanation as a final product: a statement or set of facts describing the causality of an event. However, it is important to note that, as stated by Lombrozo \cite{lombrozo:structure:2006}, an explanation is both a product and a \textit{process}. Miller argues that there are indeed two processes in an explanation, both involving human mental reasoning:

\begin{enumerate}
    \item \textit{Social Process}: The explainer exchanges information with the explainee such that the explainee has enough information to understand the causes of an event.
    \item \textit{Cognitive Process}: The explainee uses contrastive reasoning to solve what is the best cause of an event based on the information provided by the explainer.
\end{enumerate}

Because an explanation is a social process, it is ultimately upon the explainee to decide whether an explanation is sound or sufficient for understanding. Still, there are some principles which can aid in the process of human understanding when building an explanation.  Firstly, an explanation, being a social process, is ultimately a conversation (as proposed by Hilton in 1990 \cite{hilton:conversational:1990}). The difference between causal attribution and explanation, according to Hilton, is that an explanation requires an exchange between parties. The conversational model of explanation emphasizes this fact in a way that the simple causal attribution model does not.

If an explanation is a conversation, it thus follows that it should adhere to Grice's four maxims of conversation, or the \textit{Gricean maxims} \cite{grice:logic:1975}:

\begin{enumerate}
    \item \textit{Quality:} Ensure that the explanans (answers to the question) are true, to the best of the explainer's knowledge.
    \item \textit{Quantity:} Provide the right amount of information (do not provide more or less than required).
    \item \textit{Relation:} Provide information that is relevant to the conversation and exclude irrelevant information.
    \item \textit{Manner:} Conduct the conversation in an appropriate way, being easy to understand: in other words, do not be obscure or ambiguous, be orderly, and avoid an excess of words.
\end{enumerate}

One way to ensure that the maxim of \textit{Quantity} and \textit{Relation} is met is to focus on the next insight: explanations are \textit{contrastive}. They are answered not as all the de facto causes of an event (this would include too much irrelevant information) but rather in light of a counterfactual case.

\subsection{Contrastive Explanations}

As Hilton states \cite{hilton:conversational:1990},

\begin{quote}
    The key insight is to recognize that one does not explain events per se, but that one explains why the puzzling event occurred in the target case but not in some counterfactual contrast case. Explanations must therefore be relevant to some question and explain the difference between the present case and some contrast case implied by the question. Every \textit{why} question thus has an implicit \textit{rather than} built into it.
\end{quote}

With contrastive reasoning, each why-question is associated with an unspoken \textit{counterfactual}---a statement declaring the opposite of the event presupposed in the question. For example, when asking why a vehicle drove down road A, the counterfactual would be that the vehicle did not drive down road A.  The implied question, then, is, ``Why did the vehicle drive down road A \textit{instead of taking a different route} (i.e., not driving down road A)?''. Now, let us presume that the counterfactual, rather than the event, is true, and the vehicle did not drive down road A.  It may be that the vehicle's route took a longer amount of time that it would have if it had driven via road A; we now can see that the vehicle originally drove down road A in order to save time.

This, then, is the crux of contrastive reasoning: by exploring what would have happened had the \textit{rather than} scenario in any why-question been true, it is possible to uncover undesirable events which can explain why such a scenario was initially avoided.







\section{Implementation: Hospital Scheduling Domain}

We have outlined a framework for building explanations, highlighting the social, conversational, and contrastive nature which explanations should employ.  In order to implement this framework for policy-aware autonomous agents, we have built a program which outputs explanations for an autonomous agent in a natural language based on information gained from counterfactual cases.\footnote{The code for this program can be found at https://github.com/CallunaZ/Explainability_Framework_Code.}  The domain for our intelligent agent, a nurse scheduling system, is described in this section and is inspired by work on nurse scheduling by Nabeshima et al.\ \cite{nabeshima:asp-based:2025}.  The primary goal of this program is to schedule the required number of nurses for each shift in a given planning period while also adhering to nurse preferences for shift types or desired weekly days off.

\subsection{Rules and Actions}

The primary action which the agent performs is that of assigning or not assigning a nurse a certain shift on each day in the planning period.  This is accomplished through the disjunctive rule,

\begin{center}
\texttt{assigned(N, D, S) | -assigned(N, D, S) :-
    nurse(N), day(D), shift(S), not off\_day(S).}
\end{center}

Through this rule, there is either an assignment or an explicit non-assignment for each nurse, day in the planning period, and shift.  Further rules and policies narrow down which shifts are allowed to and/or should be assigned to each nurse on which day.\\

Another action the agent adheres to is that of staffing.  The staffing action counts the number of nurses that are assigned to each shift, encoded by the rule,

\begin{center}
\texttt{staffing(C, D, S) :- \#count\{ N : assigned(N, D, S) \} = C, day(D), shift(S).}
\end{center}

This action allows one to keep track of whether the lower bound of required nurses has been met for each shift.  It does not by itself enforce minimum staffing needs; this will be done by a policy listed below.  However, there is a separate hard-coded constraint which prevents the staffing on any given shift from being zero:

\begin{center}
\texttt{:- staffing(0, D, S).}
\end{center}

Additional rules assist in ensuring the program operates as intended, such as a rule stating that nurses cannot be scheduled for less than 20 work days per planning period or that a nurse cannot be assigned two different shifts on the same day.

Two final actions the agent can take are: \texttt{nurse\_fwdo} and \texttt{nurse\_swdo}, which assigns a nurse his/her first week day off and his/her second week day off, respectively.  Each week day off is chosen as a number between 0 and 6, as listed above, such that the statement \texttt{nurse\_fwdo(beth, 0)} indicates that Beth's first week day off is a Monday.

Similar to the action which assigns or does not assign a working shift to each nurse on each day, the nurses' first and second week days off are chosen through a disjunctive rule.  The rules that determine a nurse's first and second week days off are as follows:

\begin{center}
\texttt{nurse\_fwdo(N, W) | -nurse\_fwdo(N, W) :- nurse(N), weekday(W).} \\
\texttt{nurse\_swdo(N, W) | -nurse\_swdo(N, W) :- nurse(N), weekday(W).} 
\end{center}

When a nurse is scheduled a weekly day off, the action \texttt{assigned} is used to assign the \texttt{weekly\_day\_off} shift to the nurse on each day which corresponds to the aforementioned weekday.  This therefore causes the program to display both work shifts and rest shifts for each nurse throughout the planning period.

\subsection{Policies}

A total of twenty policies govern the hospital scheduling domain. The first two of these policies outline general rules that the scheduling system should follow regarding staffing weekly-day-off procedures.  The remaining 18 policies (which will not be listed due to space limitations) define specific nurse preferences the program is intended to implement. The first two policies are stated in plain English below:

\begin{enumerate}
    \item The agent is not permitted to understaff any shift.
    \item The agent is obligated to schedule a nurse's weekly days off consecutively to provide adequate rest time. (This rule is defeasible).
\end{enumerate}

The $\mathscr{AOPL}$ encoding of these two policies is listed as follows ($sr1$ indicates ``staffing rule 1'' and $dor1$ indicates ``day off rule 1''):

\begin{enumerate}
    \item $sr1(C, D, S, LB): \neg permitted(staffing(C, D, S))$ \textbf{if} $day(D),\ shift(S),\ shift\_lb(S,\ LB),$ $lt(C,\ LB).$
    \item $dor1(N, W1, W2):$ normally $obl(nurse\_swdo(N, W2))$ \textbf{if} $nurse\_fwdo(N,\ W1),$\\ $oneDayApart(W1,\ W2).$
\end{enumerate}

The penalty for a staffing violation is 5 plus the difference between the lower bound of required nurses and the actual number of scheduled nurses; e.g., if the program scheduled only 1 nurse for the night shift, which requires at least 3 nurses, the penalty would be 7: 5 + (3 - 1). This ensures that the penalty is higher the more a shift is understaffed. The penalty for a violation of consecutive days off is 3 points.  This point value was chosen to be lower than the penalty for a staffing violation because giving a nurse nonconsecutive days off would not threaten hospital operations as much as staffing a shift lower than needed. Still, in order to respect the nurse's needs as much as possible, the penalty sits at three points rather than one or two.  The penalty for the remaining policies, all describing nurse preferences, is also three points for the same reason.

These policies are encoded into ASP using multiple rules according to the $\mathscr{AOPL}$ to ASP translation described in \cite{inclezan:ASP:2023} and \cite{TUMMALA:INCLEZAN:2026}: one rule identifies the head of the rule (the action before the \textbf{if} statement), other rules define each member of the body (every relevant condition listed after the \textbf{if} statements, and still other rules define the label of the rule, the rule type (strict or defeasible), and the penalty amount for violating the rule.  Dividing the policy into several rules lets the ASP code process each component more straightforwardly, making it possible to identify when the policy is being violated. 

The code which identifies a policy violation for the first policy is written as follows, using the \texttt{add\_penalty} predicate:

\begin{quote}
  \texttt{add\_penalty(R, P) :- holds(R),} \\
  \hspace*{2cm}\texttt{head(R, -permitted(staffing(C, D, S))),} \\
  \hspace*{2cm}\texttt{staffing(C, D, S),} \\
  \hspace*{2cm}\texttt{penalty(R, P).}
\end{quote}

The logic for this code states that the agent should add a penalty $P$ if rule $R$ holds, the head of rule $R$ stated that a staffing action was not permitted, the staffing action occurred anyway, and the penalty for rule $R$ is $P$. The penalties for other policies are implemented through similar means. 

In order to enforce those policies that are applicable (i.e., are not overridden by other policies), a weak constraint is added to the program to avoid the \texttt{add_penalty} atom.

\section{Explanation Program}
This section describes the design for a program that collects counterfactual information about an agent's decision and displays an explanation for a certain question in a natural language.  The program is encoded primarily in Python, using Python's \textit{clingo} library\footnote{\texttt{https://potassco.org/clingo/python-api/5.4/}} to run ASP code and extract information accordingly.  This program is divided into \textbf{three} parts, the second two of which comprise the primary explanation-generation logic.


\subsection{Part 1: Generate Original Output}

In order to produce explanations for a given intelligent agent's behavior, the intelligent agent must produce such behavior in the first place.  While the intelligent agent's decision-making process occurs through ASP alone, the Python \textit{clingo} API is utilized in order for the ASP output to be compatible with certain procedural elements that will be described below.

\begin{figure}
    \centering
    \includegraphics[alt={A calendar displaying the hospital schedule for nurse Zelda.}, width=0.75\linewidth]{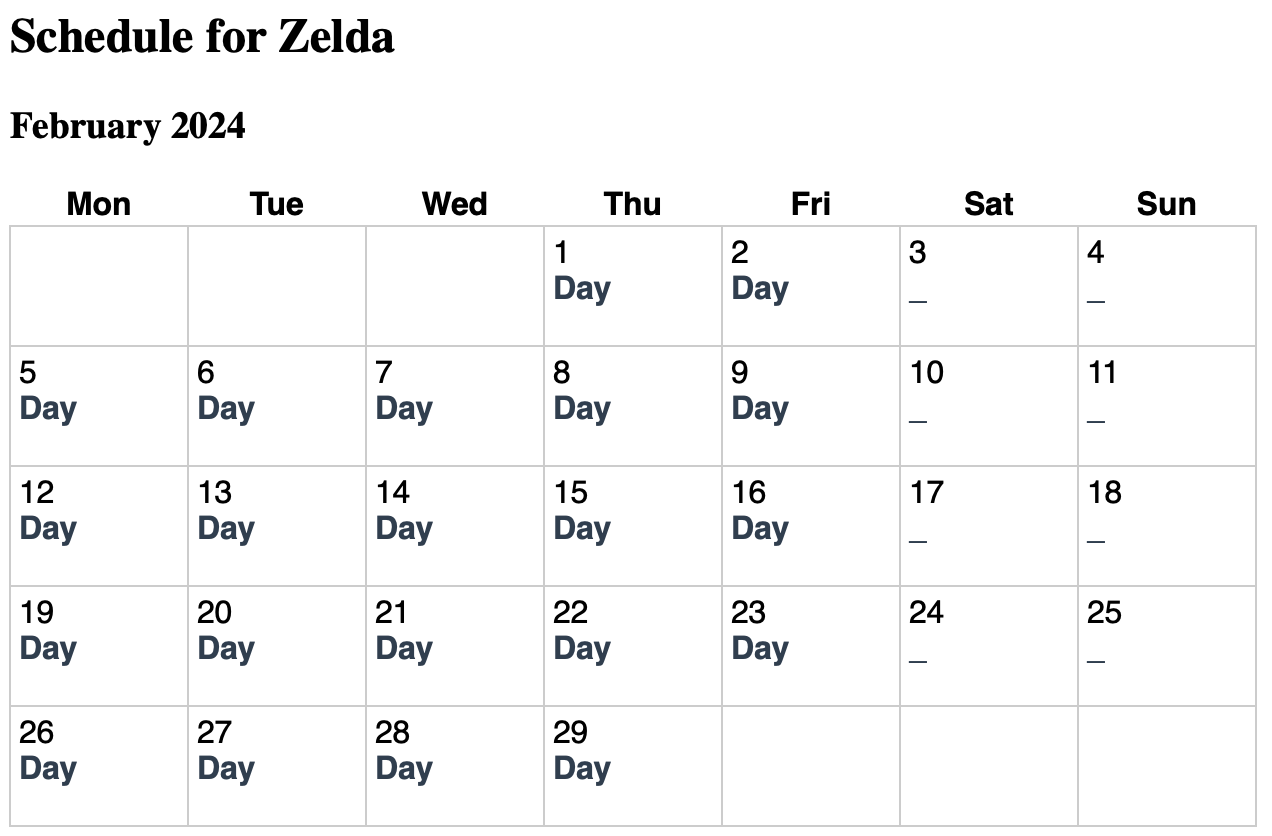}
    \caption{Nurse Zelda's schedule.}
    \label{fig:zelda-schedule}
\end{figure}

Figure \ref{fig:zelda-schedule} displays an HTML schedule for nurse Zelda as outputted by our program.  There are two policies regarding Zelda's preferences---namely, that Zelda prefers the day shift, and that Zelda prefers her first week day off to be a Saturday.  Because the optimal model does not need to violate any policies, the output correctly shows a schedule where Zelda is always assigned the day shift and where her weekly days off occur on Saturday and Sunday.

\begin{figure}
    \centering
    \includegraphics[alt={A calendar displaying the hospital schedule for nurse Howard.}, width=0.75\linewidth]{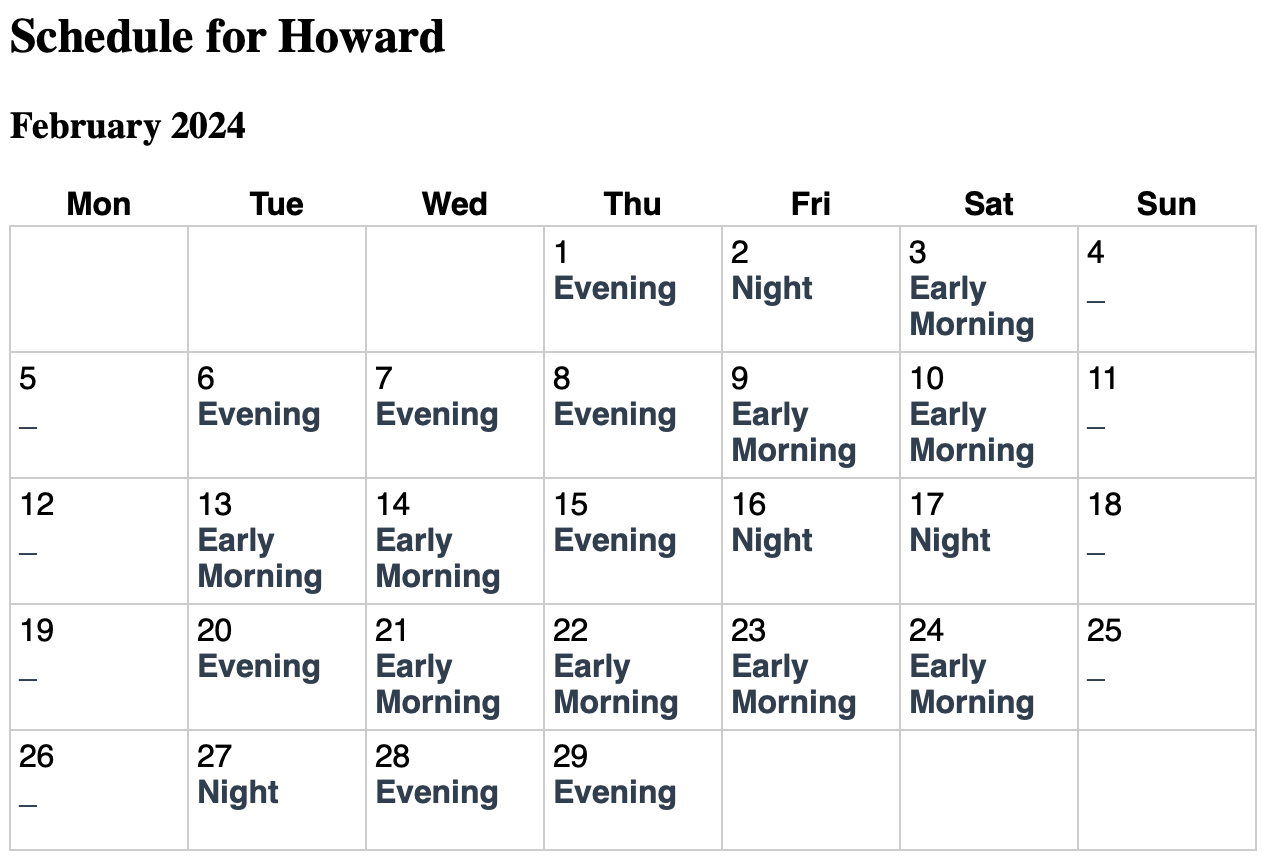}
    \caption{Nurse Howard's schedule.}
    \label{fig:howard-schedule}
\end{figure}

Figure \ref{fig:howard-schedule} displays the schedule for nurse Howard as outputted by our program.  As there are no policies dictating Howard's preferences, Howard's schedule is much more diverse than that of Zelda.  His schedule is shown to contain early morning, evening, and night shifts, along with weekly days off on Sunday and Monday.

\subsection{Part 2: Generate Questions and Counterfactuals}

The next step in our explanation program is to engage in the process of contrastive reasoning by creating counterfactual worlds. To do this, the user first needs to ask a question about one of the actions in the agent's initial output. This is accomplished by receiving input from a template of possible questions, generated through Python's TKinter library\footnote{The Tkinter library is a popular GUI library for Python.  Documentation can be found at \texttt{https://docs.python.org/3/library/tk.html}}. 

\begin{figure}
    \centering
    \includegraphics[alt={Screenshots of the question generation template for the hospital scheduling agent}, width=1.0\linewidth]{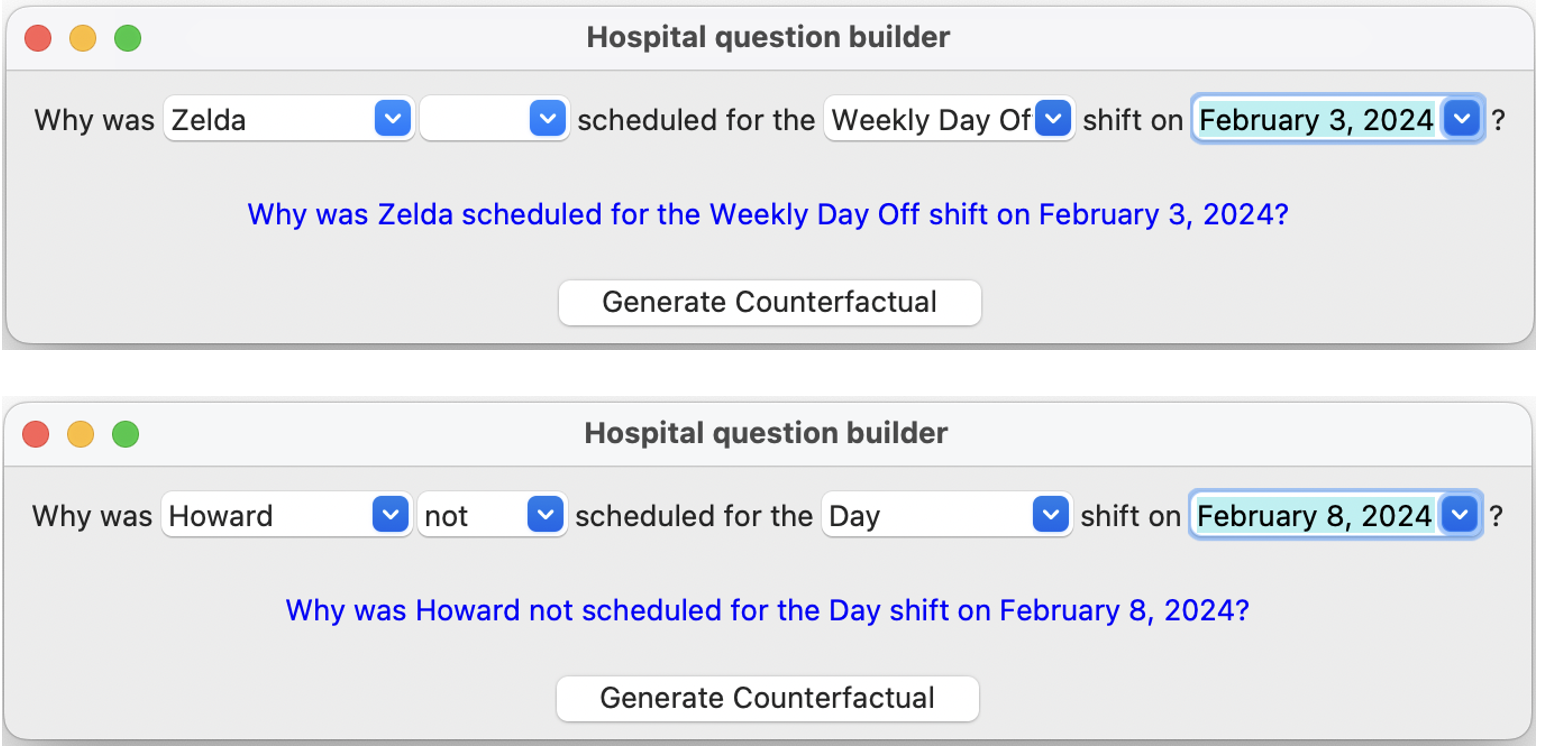}
    \caption{Hospital scheduling agent question template in action.}
    \label{fig:hospital-template}
\end{figure}

Figure \ref{fig:hospital-template} displays the template with two example questions being inputted.  These questions are,

\begin{enumerate}
    \item{Why was Zelda scheduled for the Weekly Day Off shift on February 3, 2024?}
    \item{Why was Howard not scheduled for the Day shift on February 8, 2024?}
\end{enumerate}

When the question input is submitted, a series of formatted string literals takes the input and translates it to a constraint in ASP. This constraint is stored in a file known as \texttt{counterfactual.dlv}. The constraints for the above two examples respectively appear as,

\begin{enumerate}
    \item{\texttt{:- assigned(zelda, 34, weekly_day_off).}}
    \item{\texttt{:- \textbf{not} assigned(howard, 39, day).}}
\end{enumerate}

These constraints ensure that, when the program is run again along with the counterfactual file, 1) Zelda cannot be assigned the weekly day off shift on day 34 of the year (February 3, 2024), or 2) Howard \textbf{must} be assigned the day shift on day 39 of the year (Febrary 8, 2024)---or, more specifically, the agent is not allowed to produce a model in which Howard is not assigned the day shift on day 39. Now, it is possible to see what undesirable effects, if any, shall occur when this counterfactual input is run alongside the original ASP program.

\subsection{Part 3: Generate Explanations}
The program then uses the information from the new, counterfactual world to derive explanations for the questions asked in part 2 of the program. First, let us observe what happens for the first example when Zelda is not allowed to be scheduled the weekly day off shift on February 3, 2024.  When observing the raw ASP output, one can see the atom, 

\begin{center}
    \texttt{add\_penalty(pdo4,3)}
\end{center}

This atom tells us that a penalty of 3 was added for violating the ``$pdo4$,'' or ``preferred-day-off 4'' rule. Looking at the ASP-encoded policies reveals that ``$pdo4$'' refers to the policy regarding Zelda's preferred weekly day off.  In order to read the policy and assist with the natural-language creation of explanations, an additional rule has been added to the ASP encoding of the $\mathscr{AOPL}$ rule:

\begin{center}
    \texttt{text(pdo4, "The agent must schedule Zelda's first day off on a Saturday.") :- add_penalty(pdo4, 3).}
\end{center}

This rule causes the \texttt{text} atom to appear when the penalty appears, thus enabling us to use Python's \textit{clingo} functions to iterate through and extract this text information. Using this text and additional formatted string literals, the final explanation output which would be shown to the user reads,

\begin{quote}
If Zelda were not assigned the Weekly Day Off shift on February 3, 2024, all possible schedules would violate policies which the agent is trying to adhere to. Policies include:\\
1. The agent must schedule Zelda's first day off on a Saturday. (This nurse would be scheduled a different day off, violating this policy.)
\end{quote}

This explanation tells a user why the agent scheduled Zelda the weekly day off shift on February 3, 2024: if it had not performed this action, it would have violated a policy regarding Zelda's preferred weekly day off!\\

Now, let us observe what happens in the second counterfactual world, when the agent is forced to assign Howard the day shift on February 8, 2024. In Zelda's case, the solver was easily able to refer to Zelda's specific preference to build a counterfactual model; however, for Howard's case, the solver must search through every possible nurse preference to determine whether Howard is able to ``fit'' in the schedule without violating any policies.  Thus, the \textit{clingo} solver becomes trapped in an exhaustive search space, unable to find an optimal model in a definite amount of time. Nonetheless, it is possible to leverage the handle.\textbf{wait}() function in the Python \textit{clingo} module (where ``handle'' is the clingo \textbf{SolveHandle} object) to set a timeout for the solver search.  If handle.\textbf{wait}(timeout\_seconds) returns True within ``\texttt{timeout\_seconds}'' amount of time (which we set as 5 seconds), it is safe to resume the handle and continue searching for an optimal model.  If it returns False, then no model was produced within the time limit and it is safe to assume the program is stalling.  The \textbf{SolveHandle} is canceled and the latest model which was produced can be used as the ``optimal'' model. This solution works because it realistically represents how a manager would generate a schedule in the event of an indefinite stall.

Thus, we are able to produce a counterfactual model in which Howard \textit{must} be scheduled on the Day shift on February 8, 2024.  In this world, a penalty is incurred which, in ASP, states,

\begin{center}
    \texttt{add\_penalty(ps4(39,night),3)}
\end{center}

Because this penalty contains arguments within the ``ps4(39, night)'' term, we are able to capture much more information which can be stored in a Python dictionary.  The label ``ps4'' indicates that this penalty applies to the policy for Yanni's preferred shift; the \texttt{text} rule associated with this penalty states,

\begin{center}
\texttt{text(ps4(D, S), "The agent must assign Yanni her preferred shift (day).") :- add\_penalty(ps4(D, S), 3).}
\end{center} 

The first argument inside the ``ps4'' atom, ``39,'' refers to the day on which the policy is violated: February 8, 2024.  The second argument indicates what shift Yanni was assigned instead of her preferred shift: Yanni was assigned the \textbf{night} shift.  The last argument, inside the main \texttt{add\_penalty} atom, indicates that the penalty is worth 3 points. All of the information inside this \texttt{add\_penalty} atom can be stored in a Python dictionary and retrieved when generating clarifying information about exactly what happened to violate Yanni's preferred shift policy.

Finally, because the optimal model was procured from a timeout rather than from the solver exhausting every possible solution, an additional note is added at the end of the explanation clarifying to the user that other models, likely with an equal penalty incurred, are possible (in other words, choosing to violate Yanni's shift preference was arbitrary). Thus, the output from the explanation program officially states,

\begin{quote}
If Howard were assigned the Day shift on February 8, 2024 all possible schedules would violate policies which the agent is trying to adhere to. Policies include:\\
1. The agent must assign Yanni her preferred shift (day). (This nurse would be scheduled the night shift on February 08, 2024, violating this policy.)\\

Note: There may be other possible schedules which violate different policies.
\end{quote}

Hence, this research demonstrates a framework for constructing natural-language explanations for the actions of policy-aware autonomous agents: first, a user asks a question from a template about one of the agent's actions; secondly, a counterfactual statement is generated in ASP based on the question; thirdly, the counterfactual world is generated and policy violations are identified using the \texttt{add_penalty} atom; and fourthly, those penalties are shown to the user as reasons why the agent originally performed the action in question. This framework was also tested on a planning domain \cite{TUMMALA:INCLEZAN:2026}.

\section{Evaluation Survey}
In order to understand whether the explanations outputted by the program are comprehensible to real humans, a survey was conducted with 12 human participants showing them the examples which have been outlined above. The results of this feedback are shown below. Example 1 pertains to the explanation about Zelda while Example 2 pertains to the explanation about Howard.

\subsection{Comprehensibility}

For each example scenario in the survey, the first question asks, ``Does this explanation make sense?''  There are three possible answers to this question: ``Yes,'' ``Yes, but I had to reread it in order to understand it,'' and ``No.'' Table~\ref{tbl:Make-sense} shows the percentage of people who provided each answer to this question for each example.

\begin{table}[ht]
\centering
\begin{tabular}{|l||c|c|c|}
\hline
\textbf{Example} & \textbf{Yes} & \textbf{Yes, but I had to reread it} & \textbf{No} \\ \hline\hline
Example 1 & 75\% & 16.7\% & 8.3\%\\ \hline
Example 2 & 75\% & 25\% & 0\%\\ \hline
\end{tabular}
\caption{Results for the question: ``Does this explanation make sense?''}
\label{tbl:Make-sense}
\end{table}

For each example shown above, the majority of people responded that yes, the explanation made sense. Very few people indicated that our explanations did not make sense at all.  Thus, these results indicate that our explanations make sense from the outset a majority of the time, though there is some suggestion that a few people may require more time to reread the explanations. 

\subsection{Detailed Comprehensibility}

The second question for each explanation example asks, ``How well do you understand this explanation?''  Participants answer this question on a scale of 1 to 5, with 1 being ``Not at all'' and 5 being ``Very well.'' Table~\ref{tbl:How-well} displays, for each example explanation, the percentage of people who chose each number.

\begin{table}[ht]
\centering
\begin{tabular}{|l||c|c|c|c|c|}
\hline
\textbf{Example} & \textbf{1 (Not at all)}& \textbf{2}& \textbf{3} &\textbf{4}&\textbf{5 (Very Well)}\\ \hline\hline
Example 1 & 0\% & 8.3\% & 0\% & 33.3\% & 58.3\% \\ \hline
Example 2 & 0\% & 0\% & 16.7\% & 41.7\% & 41.7\% \\ \hline
\end{tabular}
\caption{Results for the question: ``How well do you understand this explanation?''}
\label{tbl:How-well}
\end{table}

From these results, we can see that most participants responded with a 4 or a 5 for all of the example questions, indicating that the explanations could be understood quite well.

\subsection{Information Quantity}

The next question for each example asks, ``Did the explanation provide enough information to understand the agent's behavior?'' This question pertains to Grice's conversational maxim of \textit{quantity}. The exact answer choices are as follows:

\begin{itemize}
    \item No, the agent needs to provide more information.
    \item Yes, the explanation includes just enough information.
    \item The explanation provided more information than necessary.
\end{itemize}

Table~\ref{tbl:Info-quantity} displays the percentage of people that answered each question accordingly.

\begin{table}[ht]
\centering
\begin{tabular}{|l||c|c|c|}
\hline
\textbf{Example} & \textbf{Not enough} & \textbf{Right amount} & \textbf{Too much} \\ \hline\hline
Example 1 & 25\% & 66.7\% & 8.3\%\\ \hline
Example 2 & 41.7\% & 58.3\% & 0\%\\ \hline
\end{tabular}
\caption{Results for the question: ``Did the explanation provide enough information to understand the agent's behavior?''}
\label{tbl:Info-quantity}
\end{table}

These results indicated that the majority of people considered the agent to have provided the right amount of information, although Example 2 (regarding Howard) demonstrated a higher count of people who wanted more information.  The note at the end of the explanation stating that ``There may be other possible schedules which violate different policies'' may not have been as clarifying as originally intended, as one participant wrote, ``Rather than having the Note at the bottom of the explanation it should include all reasonings for the decision.''

\section{Conclusion}
To conclude, we have constructed a program which implements an explainability framework based on contrastive logic, which the social sciences have shown to be an effective tool for creating explanations.  Future work could expand upon this research by allowing for multiple lines of questioning per explanation, whether by incorporating multiple actions into one query or by allowing a back-and-forth conversation between the explanation program and the explainee.

\nocite{*}
\bibliographystyle{eptcs}
\bibliography{ThesisSourcesLogicProgramming}
\end{document}